\documentclass[jmp,author-numerical]{revtex4-1}
\usepackage{graphicx}
\usepackage{bm} 
\usepackage{epsfig}
\begin{document}

\title{Understanding looping kinetics of a long polymer molecule in solution. Exact solution for delta function sink model}

\author{Moumita Ganguly}
\email{mouganguly09@gmail.com}
\address{Indian Institute of Technology Mandi, Kamand, Himachal Pradesh - 175005, India.}

\author{Anirudhha Chakraborty}
\address{Indian Institute of Technology Mandi, Kamand, Himachal Pradesh - 175005, India.}


\begin{abstract} 
A diffusion theory for intramolecular reactions of polymer chain in dilute solution is formulated. We give a detailed analytical expression for calculation of rate of polymer looping in solution. The physical problem of looping can be modeled mathematically with the use of a Smoluchowski-like equation with a Dirac delta function sink of finite strength. The solution of this equation is expressed in terms of Laplace Transform of the Green's function for end-to-end motion of the polymer in absence of the sink. We have defined two different rate constants, the long term rate constant and the short term rate constant. The short term rate constant and long term rate constant varies with several parameters such as length of the polymer (N), bond length (b) and the relaxation time ${\tau_R}$. The long term rate constant is independent of the initial probability distribution.
\end{abstract}

\maketitle




\section{Introduction}

\noindent
Understanding the kinetics of loop formation of long chain polymer molecules has been an interesting research field both, to experimentalists \cite{Winnik,Haung,Lapidus,Hudgins} and theoreticians \cite {Wilemski,Doi,Szabo,Pastor,Portman,Sokolov,Toan}. Loop formation is believed to be an initial step in understanding several protein \cite {Hudgins,Buscaqlia} and RNA folding \cite {Thirumalai} events. The theories of loop formation dynamics are in general approximate \cite{Wilemski,Szabo}. In this paper, the looping dynamics of a single polymer chain having reactive end-groups have been modeled following the work of Szabo {\it et. al.,} \cite{Szabo}. In our model dynamics of end-to end distance is mathematically represented by a Smoluchowski-like equation for a single particle under harmonic potential in presence of a Dirac delta function sink of arbitrary strength. We have also incorporated the effect due to rate of all other chemical reactions on rate of end-to-end loop formation. The long term rate constant is shown to exhibit Arrhenius-type activation.

\section{End-to-end motion of one-dimensional polymer}
\noindent In this section we use the most simplest one dimensional description of a polymer as given by Szabo {\it et.al.,} \cite{Szabo}. Our description consists of a total $2N$ segments of unit length. The ideal chain consists of 2N monomers following random walk. Thus each monomer is allowed to take two possible orientations, one along the right and the other among the left direction. So the polymer can have any of $2^{2N}$ different conformations. In the following, we define $x$ as the end-to-end distance, with the value $x=2j$. After N steps the polymer segments can be either on the right $N+j$ or on the left $N-j$. Thus the polymer looping problem can be solved using standard machinery of probability theory. The equilibrium end-to-end distribution $P_{0j}$ of this long polymer is given by
\begin{equation}
P_{0j}=2^{-2N}(^{2N}_{N+j})
\end{equation}
Now if we consider unbiased random walk problem, then the probability distribution can be calculated by knowing the rate of fluctuations between the polymers moving to either right or left. If we imagine the polymer molecule to be immersed in solvent, the motion of polymer will be determined by various intra-molecular and intermolecular forces between the polymer and the solvent. But if we watch only the motion of polymer molecule and not the solvent, the motion would appear to be random. Then considering the variation of all right and left monomer segments being independent of each other the fluctuation is directed by the following rate equation.
\begin{equation}
\label{2}\frac{d}{dt}\left[^p_n\right]=\frac{1}{\tau_R}\left[
\begin{array}{cc}
-1&1\\
1&1
\end{array}
\right]\left[^p_n\right],
\end{equation}
where the vector $[^p_n]$ represents the activity of right and left segments orientations. $\tau_R$ is the relaxation time from one to another configuration. Now,the entire event of end-to-end looping of a polymer molecule in a solution, can be brought down to a simple random walk model confined in $2^{2N}$-dimensional configuration space. The individual monomer's reorientation would result in $2N$ ways to reorder a $x=2j$ conformation either to a $x=2j+2$ or $x=2j-2$ conformation and $N-J+1$ ways to reorient a $x=2j+2$ conformation into a $x=2j$ conformation.
Then the resulting master equation for the end-to-end distribution P(j,t)in the (2N+1)-dimensional space is given by \cite{Szabo}
\begin{equation}
\tau_R\frac{d}{dt}P(j,t)=-2NP(j,t)+(N+j+1)P(j+1,t)+(N-j+1)P(j-1,t).
\end{equation}
\noindent As we see that for long chain molecule ($N$ large), what we try to search for and measure is a distribution, {\it{i.e.}} the probability for finding the end-to-end distance of a long polymer molecule. The equilibrium distribution of  Eq.(1) may be approximated by the continuous Gaussian distribution ($x = 2 b j$)
\begin{equation}
\label{de}
P_{0}(x)=\frac{e^{-\frac{x^2}{4 b^2 N}}}{(4 \pi b^2 N)^{1/2}}
\end{equation}
Now if the individual monomers are further reduced to close points, the polymer can be represented in a continuum limit which results in Eq. (\ref{de}) as its equilibrium distribution. Then the corresponding probability conservation equation is given below
\begin{equation}
\tau_{R}\frac{\partial P(x,t)}{\partial t} = \left(4 N b^2 \frac{\partial^2}{\partial x^2} + 2 \frac{\partial}{\partial x} x \right) P(x,t).
\end{equation}
In the above `$b$' denotes the bond length of the polymer and $x$ denotes the end-to-end distance.

\section{End-to-end reaction of one-dimensional polymer}

\noindent  Now if the two ends of the polymer molecule meet, a loop would be form, {\it i.e.} at $ x = 0$. The occurrence of the looping reaction can be incorporated in our calculation by adding a $x$ dependent sink term $S(x)$ (taken to be normalized {\it i.e.} $\int_{-\infty}^{\infty} S(x)dx = 1 $) in the above equation to get
\begin{equation}
\tau_{R}\frac{\partial P(x,t)}{\partial t} = \left(4 N b^2 \frac{\partial^2}{\partial x^2} + 2 \frac{\partial}{\partial x} x  - S(x)- k_{s}\right) P(x,t).
\end{equation}
The above equation can be used to describe the dynamics of loop formation. This equation is very similar to the modified Smoluchowski equation. The term $P(x,t)$ describes the probability density of end-to-end distance at time $t$. The term $S(x)$ is a function representing sink. In our model, the effect of all other chemical reactions (involving at least one of the end group) apart from the end-to-end loop formation are incorporated through $k_{s} P(x,t)$ term.
\begin{equation}
\tau_{R}\frac{\partial P(x,t)}{\partial t} = \left(4 N b^2 \frac{\partial^2}{\partial x^2} + 2 \frac{\partial}{\partial x} x - k_{s} \right) P(x,t).
\end{equation}
The rate $k_{s}$ may be regarded as the rate of loss of probability of end-to-end distance.

\section{Exact result}
\noindent Now we do the Laplace transform $\tilde P(x,s)$ of $P(x,t)$ by \\
\begin{equation}
\tilde P(x,s)= \int^\infty_0 P(x,t) e^{-st} dt.
\end{equation}
Laplace transformation of Eq.(7) gives\\
\begin{equation}
\left[s  - {\cal L}+ \frac{1}{\tau_{R}} S(x)+\frac{k_{s}}{\tau_{R}}\right] {\tilde P}(x,s)=  P(x,0).
\end{equation}
In the above equation ${\cal L}$ is defined as follows
\begin{equation}
{\cal L} = \frac{4 N b^2}{\tau_{R}} \frac{\partial^2}{\partial x^2} + \frac{2}{\tau_{R}} \frac{\partial}{\partial x} x.
\end{equation}
The result of this equation in terms of Green's function $G(x,s|x_0)$ is
\begin{equation}
\tilde P(x,s)= \int^\infty_{-\infty} dx_{0}G(x,s+\frac{k_{s}}{\tau_{R}}|x_0)P(x_0,0),
\end{equation}
The expression $G(x,s|x_0)$ obeys the following equation
\begin{equation}
\left[s - {\cal L} + \frac{1}{\tau_{R}}S(x)\right]G(x,s|x_0)=\delta(x-x_0).
\end{equation}
Using the operator representations of quantum mechanics, we can rewrite
\begin{equation}
G(x,s|x_0)=\langle x|[s - {\cal L}+ \frac{1}{\tau_{R}}S]^{-1}|x_0\rangle,
\end{equation}
where $|x\rangle (|x_0\rangle)$ implies the position eigen-ket. Now using the operator identity
\begin{equation}
[s - {\cal L} + \frac{1}{\tau_{R}} S]^{-1}= [s - {\cal L}]^{-1}- [s - {\cal L} ]^{-1} \frac{1}{\tau_{R}} S[s - {\cal L}+ \frac{1}{\tau_{R}}S]^{-1},
\end{equation}
to obtain
\begin{equation}
G(x,s|x_0)=\langle x|[x - {\cal L}]^{-1}|x_0\rangle -\langle x|[s - {\cal L}]^{-1}\frac{1}{\tau_{R}}S[s - {\cal L} + \frac{1}{\tau_{R}} S[s - {\cal L} + \frac{1}{\tau_{R}} S]^{-1}|x_0\rangle.
\end{equation}
Inserting the intent of identity I= $\int^\infty_{-\infty} dy |y \rangle \langle y|$ in the appropriate places in the second term of the above equation, we get the following equation. 
\begin{equation}
G(x,s|x_0)= G_{0}(x,s|x_0) - \int^{\infty}_{-\infty} dy G_{0}(x,s|y)S(y)G(y,s|x_0)
\end{equation}
\noindent $G_{0}(x,s|x_0)$ is defined by
\begin{equation}
G_{0}(x,s|x_0)=\langle x|[x - {\cal L}]^{-1}|x_0\rangle
\end{equation}
and corresponds to change in end-to-end distance of the polymer, that has the inceptive value $x_0$, in the absence of any sink. It is noteworthy that Laplace transform of $G_0(x,t|x_0)$ gives the probability that the end-to-end distance of a polymer per say, starting at $x_0$ may be found at $x$, at time $t$. It obeys the following equation,
\begin{equation}
\left[(\partial /\partial t)- {\cal L} \right] g_0(x,t|x_0)=\delta(x-x_0).
\end{equation}
The above equation doesn't have a sink term in it. In the absence of sink, there is no absorption of the particle. Therefore, $\int^\infty_{-\infty} dx g_0(x,t|x_0) = 1$. From this we can conclude 
\begin{equation}
\int^\infty_{-\infty} dx G_0(x,s|x_0) = 1/s
\end{equation}
If $S(y)= k_{0}\delta(y - x_s)$, then Eq.(16) becomes \\
\begin{equation}
G(x,s|x_0)=G_0(x,s|x_0) - \frac{k_0}{\tau_{R}}\;G_0(x,s|x_s)G(x_s,s|x_0).
\end{equation}
We now solve Eq.(20) to find
\begin{equation}
G(x_s,\;s|x_0)=G_0(x_s,\;s|x_0) [1+ \frac{k_0}{\tau_{R}} G_0(x_s,\;s|x_s)]^{-1}.
\end{equation}
When substituting the above back into Eq.(20) gives
\begin{equation}
G(x,s|x_0)=G_0(x,s|x_0)- \frac{k_0}{\tau_{R}} G_0(x,s|x_s)G_0(x_s,s|x_0)[1+ \frac{k_0}{\tau_{R}} \;G_0(x_s,s|x_s)]^{-1}.
\end{equation}
Using the expression of $G(x,s|x_0)$ in Eq.(11) we get $\tilde P(x,s)$ explicitly. It is difficult to calculate survival probability $P_e(t) =\int^\infty_{-\infty} dx P(x,t)$. Instead one can easily calculate the Laplace transform $P_e(s)$ of $ P_e(t)$ directly. $P_e(s)$ is associated to $P(x,s)$ by 
\begin{equation}
P_e(s) = \int^\infty_{-\infty} dx {\tilde P}(x,s).
\end{equation} 
From Eq. (11), Eq. (22) and Eq. (23), we get
\begin{equation}
P_e(s)=\frac{1}{s+k_{s}}\left[1-[1+\frac{k_0}{\tau_{R}} G_0(x_s,s+k_{s}|x_s)]^{-1} \frac{k_0}{\tau_{R}} \; \times \int^\infty_{-\infty} dx_0 \; G_0 (x_s,s+k_{s}|x_0)P(x_0,0).\right]
\end{equation}
The average and long time rate constants can be derived from $P_e(s).$ Thus, $k^{-1}_I =P_e(0)$ and $k_L$ = negative of the pole of $P_e(s),$ which is close to the origin. From (23), we obtain
\begin{equation}
k^{-1}_I =\frac{1}{k_{s}}\left[1- [1+\frac{k_0}{\tau_{R}} G_0(x_s,k_{s}|x_s)]^{-1} \frac{k_0}{\tau_{R}} \; \times \int^\infty_{-\infty} dx_0 \; G_0 (x_s,k_{s}|x_0)P(x_0,0).\right]
\end{equation}
Thus $k_I$ depends on the initial probability distribution $P(x,0)$ whereas $k_L = - $ pole of $[\;[ 1+\frac{k_0}{\tau_R}\; G_0(x_s, s+k_s|x_s)][s+k_s]\;]^{-1}$, the one which is closest to the origin, on the negative $s$ - axis, and is independent of the initial distribution $P(x_0,0)$.
The $G_0(x,s;x_0)$ can be found out by using the following equation {\cite{KLS}:
\begin{equation}
\left(s - {\cal L}\right) G_{0}(x,s;x_0)= \delta (x - x_0) 
\end{equation}
Using standard method \cite{Hilbert} to obtain.
\begin{equation}
G_0(x,s;x_0)=F(z,s;z_0)/(s+k_s)
\end{equation}
with
\begin{equation}
F(z,s;z_0)= D_\nu(-z_<)D_\nu(z_>)e^{(z_0^2-z^2)/4}\Gamma(1-\nu)[1/(4 \pi N b^2)]^{1/2} 
\end{equation}
In the above, $z$ defined by $z = x(2Nb^2)^{1/2}$  and $z_j = x_j(2Nb^2)^{1/2}$, $\nu  = —s{\tau_{R}}/2$ and $\Gamma(\nu)$ is the gamma function. Also, $z_{<}= min(z, z_0)$ and $z_{>}= max(z, z_0)$. $D_{\nu}$ represent parabolic cylinder functions. To get an understanding of the behavior of $k_I$ and $k_L$, we assume the initial distribution $P^0_e(x_0)$ is represented by $\delta(x-x_0)$. Then, we get 
\begin{equation}
{k_I}^{-1}= {k_s}^{-1}\left(1 - \frac{\frac{k_0}{\tau_{R}}F(z_s,k_s|z_0}{k_s+ {\frac{k_0}{\tau_{R}}}F(z_s,k_s|z_s)} \right)
\end{equation}
Again
\begin{equation}
k_L= k_r - [ values \; of \; s \; for \; which \;\; s+ {\frac{k_0}{\tau_{R}}} F(z_s,s|z_s)=0]
\end{equation}
We should mention that $k_I$ is dependent on the initial position $x_0$ and $k_s$ whereas $k_L$ is independent of the initial position.
In the following , we assume $k_s\rightarrow$ 0, in this limit we arrive at conclusions, which we expect to be valid even when $k_s$ is finite. Using the properties of $D_v{(z)}$, we find that when $k_s\rightarrow 0, F{(z_s,k_s|z_0)}$ and $F{(z_s,k_s|z_s)}\rightarrow exp(-z_s^2/2){[1/(4\pi Nb^2)]}^\frac{1}{2}$so that
\begin{equation}
\frac{k_0}{\tau_r} F{(z_s,k_s|z_0)}/[k_s+ \frac{k_0}{\tau_r}F{(z_s,k_s|z_s)}]\rightarrow 1.
\end{equation}
\\Hence 
\begin{equation}
k_I^{-1}=-{[\frac{\partial}{\partial k_s}\left[\frac{\frac{k_0}{\tau_R} F(z_s,k_s|z_0)}{k_s + \frac{k_0}{\tau_R} F(z_s,k_s|z_s)}\right]}_{k_s \rightarrow 0}
\end{equation} 
If we take $z_0 < z_s $, so that the particle is initially placed to the left of sink. Then \\
\begin{equation}
k_I^{-1}= \frac{e^{{z_s}^2/2} \tau_R}{k_0}{[1/{(4\pi N b^2)}]}^{1/2}+ \left[\frac{\partial}{\partial k_s}\left[\frac{e^{[(z_0^2-z_s^2)/4]}D_v{(-z_0)}}{D_v{(-z_s)}}\right]\right]_{v=0}
\end{equation} 
After simplification
\begin{equation}
k_I^{-1}= \frac{e^{{z_s}^2/2} \tau_R}{k_0}{[1/{(4\pi N b^2)}]}^{1/2}+ \left(\int_{z_0}^{z_s} dz e^{(z^2/2)}\left[1+erf(z/\sqrt{2}\right]\right)(\pi/2)({\tau_R/2})
\end {equation}
The long-term rate constant $k_L$ is determined by the value of $s$, which satisfy $s+ \frac{k_0}{\tau_R}F(z_s,s|z_s)=0$. This equation can be written as an equation for $\nu (= -s{\tau_R} /2)$
\begin{equation}
\nu = D_\nu(-z_c)D_\nu(z_c)\Gamma(1-\nu)\frac{k_0}{4 b \sqrt{\pi N}} 
\end{equation}
For integer values of $\nu$, $D_\nu(z)=2^{-\nu/2}e^{-z^2/4}H_{\nu}(z/\sqrt{2})$, $H_{\nu}$ are Hermite polynomials. $\Gamma(1-\nu)$ has poles at $\nu = 1,2, . . . .$. Our interest is in $\nu \in [0, 1]$, as $k_L = \frac{2}{\tau_R} \nu$ for $k_s =0$. If $ \frac{k_0}{(4 b \sqrt{\pi N})}\ll 1$, or $z_c \gg 1$ then $\nu \ll 1$ and one can arrive
\begin{equation}
\nu = D_0(-z_s)D_0(z_s)\frac{k_0}{(4 b \sqrt{\pi N)}} 
\end{equation}
and hence 
\begin{equation}
k_L = \frac{e^{{-z_s}^2/2} \tau_R}{k_0}{[1/{(4\pi N b^2)}]}^{1/2}
\end{equation}
In this limit, the rate constant $k_L$ exhibits Arrhenius type activation.

\begin{table}[ht]
\caption{Our  Results} 
\centering 
\begin{tabular}{c c c }
\hline\hline 
 & $k_{L}$  & $k_{I}$ \\ [0.5ex] 
\hline 
 Square root of chain length ${\sqrt N}$  & Inversely Proportional & Directly proportional \\ 
 Bond length $b$        & Inversely Proportional & Directly Proportional \\
 Relaxation time ${\tau_R}$    & Directly Proportional & Inversely Proportional\\
\hline 
\end{tabular}
\label{table:nonlin} 
\end{table}

\section{Conclusions:}

\noindent In this paper, we have given a general procedure for solving the problem of looping dynamics of long polymer molecule in solution. The discrete polymer equation boils down to motion of a single particle with harmonic oscillator. Explicite expressions for $k_{I}$ and $k_{L}$ have been derived. The short term rate constant depends on the initial distribution whereas the long term rate constant is independent of the initial distribution of the polymer chain. A very unique and improved argument presented in this paper is that the rate of looping considers not only the end-to-end looping but occurrence of all other chemical reactions involving at least one of the end groups of the polymer chain. This same method can also be used for the case where sink is a non-local operator and can be written as $S = |f> k_{0} <g|$ where f and g are arbitrary functions. If one choose both of them as Gaussian functions, the result will be an interesting improvement over the current model. The sink $S(x)$ may also be taken as some appropriate linear combination of such operators.

\section{Acknowledgments:}
\noindent One of the author (M.G.) would like to thank IIT Mandi for HTRA fellowship and the other author thanks IIT mandi for providing CPDA grant.

\end{document}